\documentclass[prl,aps,a4paper,showpacs,twocolumn,floatfix]{revtex4}

\usepackage{graphicx}
\usepackage[ansinew]{inputenc}
\usepackage{array}
\usepackage{color}
\usepackage{amsmath}
\usepackage{amsxtra}
\usepackage{amstext}
\usepackage{amssymb}
\usepackage{latexsym}
\usepackage{dsfont}
\begin{document}

\title{Towards all-optical optomechanics: An optical spring mirror}

\author{S. Singh, G. A. Phelps , D. S. Goldbaum, E. M. Wright and P. Meystre}
\affiliation{B2 Institute, Department of Physics and College of Optical
Sciences\\The University of Arizona, Tucson, Arizona 85721}

\date{\today}

\begin{abstract}
The dominant hurdle to the operation of optomechanical systems in the quantum regime is the coupling of the vibrating element to a thermal reservoir via mechanical supports. Here we propose a scheme that uses an {\it optical spring} to replace the mechanical support. We show that the resolved-sideband regime of cooling can be reached in a configuration using a high-reflectivity disk mirror held by an optical tweezer as one of the end-mirrors of a Fabry-P{\' e}rot cavity. We find a final phonon occupation number of the trapped mirror ${\bar n}$= 0.14 for reasonable parameters, well within the quantum regime. This demonstrates the promise of dielectric disks attached to optical springs for the observation of quantum effects in macroscopic objects.
\end{abstract}


\maketitle

Operating macroscopic objects in the quantum regime is a challenge whose successful completion will have profound implications, ranging from an improved fundamental understanding of the quantum-classical interface and of the quantum measurement process to the development of quantum detectors of unsurpassed sensitivity. Cooling a nanomechanical system to its ground state of center-of-mass motion is an important step toward that goal, and spectacular progress has recently occurred via an interdisciplinary approach combining tools from nanoscience, quantum optics, and condensed matter physics. A recent benchmark experiment has demonstrated the operation of a micromechanical resonator down to a phonon number ${\bar n}< 0.07$, as well as quantum control at the single-phonon level~\cite{OConnell2010}. Such developments open up the way to the detection of exceedingly feeble forces and displacements, with applications ranging from the quantum control of molecular processes to gravitational wave detection \cite{KippenbergScience09}.

One of the simplest systems being considered in this quest consists of a small vibrating element that forms one of the end-mirrors of a Fabry-P\'erot cavity \cite{Meystre1985}. So far the biggest hurdle in achieving the ground-state cooling of such a mirror has been the coupling to a thermal reservoir by way of a mechanical support. This support acts as the dominant source of dissipation and decoherence. The goal of this note is to theoretically discuss an alternative configuration where the mechanical clamping of the system is completely eliminated, replaced by an optical spring realized by an optical tweezer.

There is a large volume of work on the trapping of dielectric particles -- from atoms to bacteria, in the focus of laser beams far detuned from any electronic resonance \cite{Ashkin1997}. Over the last two decades optical tweezers have matured into a well established tool, providing elegant and relatively simple ways to control the motion and to measure the weak forces acting on particles suspended in a fluid or in vacuum. A key observation in the present context is that macroscopic objects optically levitated in vacuum are remarkably isolated from most environmental noise sources \cite{Ashkin1976}, and as such, should provide a route toward the elimination of the clamping losses already mentioned. Exploiting this idea, several recent theoretical proposals have considered levitating macroscopic objects (spheres or even living organisms) in a cavity and cooling them to their ground state of center-of-mass motion \cite{Oriol2010,Chang2010}. Our work builds further on these ideas. As we shall see, trapping and cooling of a dielectric end mirror of a resonator, rather than an object inside a resonator, presents a number of advantages. In particular, scattering losses are significantly reduced compared to the case of spheres. Furthermore, one can envision simple schemes to couple it to a two-level atom in order to fully characterize and control the quantum state of the mechanical motion \cite{SinghPM2010}.
\begin{figure}[t]
\includegraphics[width=0.37 \textwidth]{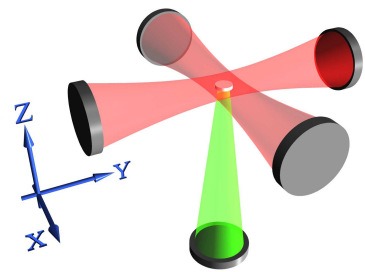}
\caption{\label{fig:setup}(Color online).
Arrangement for an optomechanical cavity without clamping losses. The disk mirror is trapped in the optical tweezer due to the crossed elliptical Gaussian beams shown in red, and provides the moving mirror for the Fabry-P\'erot aligned along the z-axis shown in green.}
\end{figure}

The optically trapped mirror that we consider is a Bragg disk composed of alternating layers of two dielectrics. It is held in vacuum by the optical gradient force due to two linearly polarized elliptical gaussian beams of equal wavelengths $\lambda$. The disk axis is along the $z$-axis of the Fabry-P\'erot interferometer, and perpendicular to the trap beams, see Fig.~1. The tweezer beam traveling in the $x$-direction is polarized along the $y$-direction, and the beam traveling in the $y$-direction is polarized in the $x$-direction; the orthogonal polarizations being chosen to avoid the onset of interferences in the overlap region of the beams. Both beams have an elliptical transverse profile with the smallest beam waist along $z$, so as to provide a tight confinement along that axis. The total intensity of the trapping beams has the form
\begin{eqnarray}
 I({\bf r})=I_{0x}\frac{\exp\left[\frac{-2y^2}{w_{0y}^2(1+x^2/y_r^2)}+\frac{-2z^2}{w_{0z}^2(1+x^2/z_r^2)}\right]}
{\sqrt{\left(1+x^2/y_r^2\right)\left(1+x^2/z_r^2\right)}}\nonumber \\
+I_{0y}\frac{\exp\left[ \frac{-2x^2}{w_{0x}^2(1+y^2/x_r^2)}+\frac{-2z^2}{w_{0z}^2(1+y^2/z_r^2)}\right]}{\sqrt{\left(1+y^2/x_r^2\right)\left(1+y^2/z_r^2\right)}},
\end{eqnarray}
where $I_{0x}$ and $I_{0y}$ are the on-axis intensities of the laser beams traveling in the $x$ and $y$ directions, $w_{0\mu}$ is the focussed beam waists with $\mu=x,y,z$, and $\mu_r=\pi w_{0\mu}^2/\lambda$ the Rayleigh ranges along the respective directions.

For concreteness we consider the case of a Nd:YAG trapping laser ($\lambda$=1.064 $\mu$m) that is far detuned from any material resonance in the dielectric disk. In this far-detuned limit we may assume that the field induces a dipole moment in the material $\bf{p}={\mathbf\alpha}{\bf E}$, where $\bf{\alpha}$ is the polarizability tensor and $\bf{E}$ the electric field envelope. Further assuming that the field envelope varies little over the dimensions of the disk, the components of the polarizability tensor can be approximated by those induced by a static electric field.  The static polarizability of a dielectric cylinder in a static field has previously been calculated numerically \cite{Venermo2005}. Instead, we use the analytical expression for the polarizability of a spheroid \cite{Landau1968}, which is close to that of a cylinder of the same permittivity $\epsilon$ and aspect ratio. For our parameters, that approximation results in an error of about 5\% in the value of the components of the polarizability tensor. The longitudinal and transverse polarizabilities of a spheroid of diameter $d$, length $l$, eccentricity $e=\sqrt{(d/l)^2-1}$ and volume $V$ are then given by
\begin{equation}
\label{eq:polarizability}
\alpha_{\perp, z}=\epsilon_0 V \left [\frac{\epsilon_r-1}{1+N_{\perp, z}(\epsilon_r-1)} \right ],
\end{equation}
where $\epsilon_r=\epsilon/\epsilon_0$ is its relative permittivity, $N_z=(1+e^2)\left(e-\arctan{e}\right)/e^3$, and $N_{\perp}=0.5 (1-N_z)$.

The Bragg disks under consideration consist of alternating layers of two materials that each have a thickness of a quarter wavelength $\lambda_{FP}/4$ of the Fabry-P\'erot laser \cite{Rempe1992, Groblacher2009}. For the purpose of estimating the properties of the trapped disk we make two approximations, namely that the disk height $h$ and the disk diameter $d$ obey $h<w_{0z},d<x_r,y_r$, so that the field varies little over the disk, and that for purposes of estimating the trapping frequencies we may replace the layered structure by a dielectric slab with an effective permittivity.  Then for a disk with $d=100~\mu$m, $h=4~\mu$m, and mass $m=1.48 \times 10^{-10}$ kg made of stacked SiO$_2$/Ta$_2$O$_5$ layers with effective permittivity $\epsilon = 5.9\epsilon_0$, we find a polarizability $\alpha_{\perp}=1.20\times 10^{-24}~{\rm C} \cdot {\rm m}^2 {\rm V}^{-1}$ and $\alpha_{z}=2.44\times 10^{-25}~{\rm C}\cdot{\rm m}^2 {\rm V}^{-1}$.

The optical potential due to the gradient force is $V({\bf r})=-\alpha_\perp I({\bf r})/(2 \epsilon_0 c)$, ${\bf r}$ being small displacements about the origin.  For small deviations along the $z$-axis this yields a harmonic potential with frequency
\begin{equation}
\label{eq:zfreq}
\omega_z=\left [\frac{2 \alpha_\perp}{m c \epsilon_0 w_{0z}^2}(I_{0x}+I_{0y})\right ]^{1/2},
\end{equation}
For Nd:YAG laser beams of intensity 80 mW/$\mu$m$^2$ and beam waists $w_{0x}=w_{0y}=200~\mu$m and $w_{0z}=8~\mu$m, we then find $\omega_z = 1.24 \times 10^5$ rad/s, and in a similar manner we find $\omega_{x,y} = 4\times10^3$ rad/s for the transverse trapping frequencies.

Next we assess the angular motion of the disk with respect to the $x$ and $y$ axes, see Fig.~1. In particular, we calculate the wobble frequency $\omega_{\rm wob}$ of the disk when it is misaligned by an angle $\theta$ with respect to the $x$-axis. Such motion of asymmetric isotropic objects in linearly polarized optical traps has previously been studied in detail, for example in Ref.~\cite{Bonin2002}. We estimate $\omega_{\rm wob}$ by considering a light beam propagating in the $y$-direction and polarized along $x$. For a disk misaligned by an angle $\theta$ with respect to the $x$-axis the induced dipole moment is $\textbf{p}=[\alpha_\perp E_0 \cos{\theta}\textbf{\^{x}}+\alpha_z E_0\sin{\theta}\textbf{\^{z}}$]. An analysis of small angle harmonic rotational motion along $y$ shows that is has the frequency
\begin{equation}
\omega_{\rm wob}=\left[\frac{12 I_{0y} (\alpha_\perp-\alpha_z)}{\epsilon_0 c {\cal I}_x}\right]^{1/2}
\end{equation}
where ${\cal I}_{x}=m(3d^2/4+h^2)$ is the moment of inertia of the disk along $x$.
For the parameters employed here we find $\omega_{\rm wob}=1.8\times10^4$ rad/s.  We note that $\omega_z>> \omega_{\rm wob}$ thereby ruling out any parametric coupling between the wobble mode and the longitudinal mirror motion.  This means that the wobble mode should not be detrimental to cooling the longitudinal mirror motion.

Although the mirror is nominally transparent to the trapping lasers, it will absorb some light, and with no heat sinking the only way to dissipate this energy is through blackbody radiation. This heating of the dielectric due to energy absorbed is described in detail in \cite{Oriol2010}. For an imaginary component of permittivity $\epsilon''=10^{-10}\epsilon_0$ \cite{Oriol2010} we find that the temperature of the mirror increases by about 2K, meaning that the heating of the mirror is not significant.

Having established the mechanical properties of the trapped Bragg disk, we now turn to a discussion of the Fabry-P{\'e}rot in which the Bragg disk serves as a vibrating end-mirror \cite{KippenbergScience09}. The fixed mirror of the Fabry-P\'erot interferometer, assumed to have a reflectivity $R_f$=0.999998, is placed at a distance $L$=15~cm from the movable mirror of lower reflectivity $R_m$=0.9998. We note that small mirrors of comparable or smaller sizes with reflectivity exceeding 0.9998 are already being used in experiments \cite{Groblacher2009}. For $\lambda = 852$ nm, the cavity damping rate is $\kappa=\pi c/\mathcal{F} L \approx 200$ kHz ($\mathcal{F}$ is the finesse), a value comparable to the optical trap frequency, so that the system is only marginally approaching the resolved side-band limit of radiation pressure cooling. Ignoring all sources of noise, these parameters result in a minimum thermal phonon occupation number of \cite{Vahala2007}
\begin{equation}
\langle n\rangle_{\rm min} = -\frac{4(\Delta+\omega_z)^2+\kappa^2}{16 \omega_z \Delta}.
\end{equation}
For our parameters and a detuning $\Delta = (\omega_{\rm laser}-\omega_{c})=-160$ kHz from the cavity resonance ($\omega_{c}$), we get $\langle n\rangle_{\rm min}\simeq 0.14$, well into the quantum regime.  We remark that we may reduce this value by using tighter trapping, though this would violate our approximation that the field varies little over the dimensions of the disk.  The quoted value is thus a lower value consistent with our approximations, but by no means a fundamental limit.

In calculating $\langle n\rangle_{\rm min}$ we have ignored all effects of noise. The major sources of noise are the fluctuations of the trapping and the Fabry-P{\'e}rot cavity lasers, and background gas collisions.  We next evaluate their impact on $\langle n\rangle_{\rm min}$.

{\it Trapping laser fluctuations:} There are three noise sources due to the optical tweezer laser beams: intensity fluctuations, beam-pointing fluctuations, and photon scattering losses. The first two noise sources have been studied extensively in the context of trapping alkali atoms in optical traps \cite{Gehm98}. The intensity fluctuations lead to a change in trap frequency, see Eq.~(\ref{eq:zfreq}), resulting in transitions $n \rightarrow n\pm 2$ between states of vibration of the trapped mirror. This produces a rate of parametric heating due to intensity fluctuations given by
\begin{equation}
\gamma_{I}= \frac{1}{4} \omega_{z}^2 S_I(2\omega_{z}),
\end{equation}
where $S_I(2\omega_z)$ is the noise power spectrum of the laser. For the concrete example that we consider $S_{I}= 10^{-12}$ Hz$^{-1}$, resulting in a heating rate of $3.8 \times 10^{-3}$/s. We note that Nd:YAG lasers with a lower noise spectrum are available and would further reduce this source of heating.

Beam-pointing fluctuations cause fluctuations of the trap center and lead to a constant heating rate given by $ {\dot\gamma}_x= \frac{1}{4} \omega_{z}^4mS_x(\omega_{z})$. For a spectrum of position fluctuations  $S_x(\omega_z)$ of $10^{-10}$ $\mu$m$^2$ Hz$^{-1}$ this yields a negligible heating rate of the order of $10^{-16}$/s.

The origin of scattering losses is the momentum noise resulting from the fluctuations of trapping laser photons impinging on the two surfaces of the disk. The resulting momentum fluctuations along $z$ are given by
$\Delta p_{\rm scat}/\Delta t=\sqrt{2n_0}\hbar k\theta_z$,
where $n_0$ is the mean flux of trapping laser photons hitting the disk , and $\theta_z$ is the angle between the disk's surface and the direction of photon momentum, which is very small for the geometry under consideration. The factor of $\sqrt{2n_0}$ stems from the Poisson statistics of the laser intensity. For our parameters and an angle of $10^{-2}$ radians, this rate is of the order of $10^{-19}$/s, and hence completely negligible, in contrast to the situation with dielectric spheres, where this scattering mechanism is the dominant source of noise \cite{Oriol2010, Chang2010}.

{\it Fabry-P\'erot laser fluctuations:} Another source of noise that places a fundamental limit on the occupation number of the center-of-mass motion of the moving mirror is the linewidth of the Fabry- P\'erot laser \cite{Rabl2009}. Here we model the laser linewidth in terms of a phase diffusion process that drives the laser field $E_{in}e^{i\phi(t)}$. Here the phase $\phi(t)$ is given by
\begin{equation}
\phi(t) = \sqrt{2 \Gamma_L} \int_0^t \eta(s) ds,
\end{equation}
where $\Gamma_L$ is the laser linewidth and $\eta(s)$ is a gaussian white noise process with mean $\langle \eta(s) \rangle = 0$ and correlation $\langle \eta(s)\eta(v) \rangle = \delta(s-v)$. For $|(\omega_{c}z)/(\omega_z L)| \ll 1$ this results in the linewidth-modified cooling rate
\begin{equation}
\gamma_{rp} = - \left(\frac{\omega_{c}\kappa}{m \omega_z L^2}\right) \frac{8P_{in}\left[A_--A_+\right]}{\left[\left(2\Gamma_L+\kappa\right)^2+4\Delta^2\right]\left(\kappa^2+\omega_z^2\right)}
\end{equation}
where $P_{in}$ is the input power and $A_\pm$ is given by
\begin{equation}
A_{\pm} = \frac{\left(\Gamma_L+\kappa\right)\left(2\Gamma_L+\kappa\right)^2+2\Gamma_L((\Delta\pm\omega_z)^2+\Delta^2)+\kappa\omega_z^2}{\left(2\Gamma_L+\kappa\right)^2+4\left(\Delta\pm\omega_z\right)^2}.
\end{equation}

\begin{figure}[t]
\includegraphics[width=0.45 \textwidth]{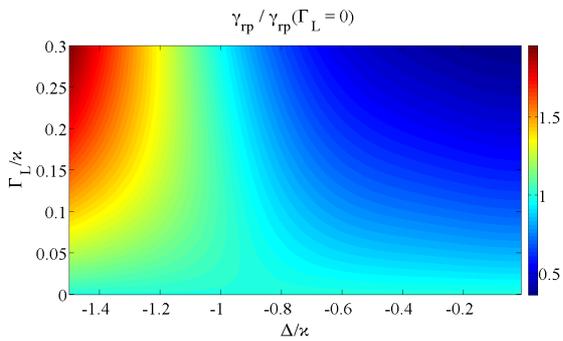}
\caption{\label{fig:noise}(Color online).
Dynamical backaction cooling rate in the presence of laser phase noise, scaled to that of a perfectly monochromatic laser, as a function of detuning and laser linewidth, normalized to $\kappa$.}
\end{figure}

The effect of the laser linewidth on the cooling rate is shown in Fig.~\ref{fig:noise}, which illustrates that it generally makes backaction cooling less efficient. However, there is a range of detunings $\Delta$ for the cooling rate is essentially unchanged from the case of a perfectly monochromatic laser ($\Gamma_L=0$), a result of the excitation of the anti-Stokes sideband from higher frequencies in the laser spectrum.

For our parameters, a 0.1mW laser of linewidth 10 kHz, detuned -160 kHz from the cavity resonance results in a cooling rate $\gamma_{rp}$ of $2.21\times10^{7}$ /s.

{\it Background gas collisions:} The fluctuations in mirror motion due to background gas collisions can be described by the Langevin equation $\ddot{z}+\gamma_{\rm bg}\dot{z}=\xi(t)$
where the fluctuating force $\xi(t)$ obeys the markovian correlation relations
$ \langle \xi(t)\rangle = 0 , \langle \xi(t)\xi(t')\rangle=q\delta(t-t')$ with $q$ given by the fluctuation-dissipation theorem as $q=2k_BT\gamma_{bg}/m$. To derive an expression for $\gamma_{bg}$ we consider motion along the z-axis only. A gas molecule of mass $m_g$ and velocity $v_g$ undergoing an elastic collision with the disk imparts a momentum change $\delta p = 2 m_g v_{g}$. In the moving frame of the disk, this gives $\Delta p_{\rm disk} = 2m_g (v_{g}-v_{disk})-2m_g(v_{g}+v_{disk})$, the two contributions corresponding to forward and backward collisions. The rate of momentum transfer is then obtained by multiplying this expression by the number of collisions per unit time ($nA v_{g}/2$), where $n$ is the number density of gas molecules, $A$ is the cross-section area of the disk, and $v_g$ is the mean speed of the molecules, taken to be to be the average thermal velocity for an ideal gas of pressure $P$. This gives
\begin{equation}
 \gamma_{bg}=\frac{4PA}{mv_{g}},
\end{equation}
For a pressure of $10^{-6}$ torr, we find $\gamma_{bg} = 5.45\times10^{-5}/$s.

Both intensity fluctuations and background collisions are mechanisms of damping for the disk mirror and provide the equivalent of a mechanical $Q$-factor. The coupling to a thermal reservoir increases the attainable mean phonon number to $\langle n\rangle_{\rm min}$ by $\gamma_m n_R/(\gamma_{rp}+\gamma_{m})$, where $\gamma_m$ is the mechanical damping, due here to $\gamma_{bg}$ and $\gamma_I$ and $n_R$ is the average occupation number of the relevant mode before cooling, $n_R \simeq k_{B}T/\hbar\omega_z$. For our parameters, the contribution of this mechanical damping is very small, $\simeq 0.05$, and can be reduced further via better stabilized lasers and an improved vacuum.

In conclusion, we have shown that the coupling to the thermal reservoir in standard optomechanical setups can be completely eliminated by optical levitation of the Fabry-P\'erot mirror. Importantly, the minimum thermal occupation number discussed in this letter is limited by our approximations, and/or technical parameters- but not by fundamental constraints. In particular, we note that $\langle n\rangle_{\rm min}$ can be further lowered by stiffening the optical spring holding the moving mirror in place. This optical spring effect has been studied intensively in the gravitational wave detection community \cite{Corbitt2007}, where the moving mirror's mechanical resonance frequency has been enhanced by a few orders of magnitude by a two-laser configuration with one of the wavelengths blue-detuned from the cavity resonance. A similar approach could also increase $\omega_z$ in our case without increasing the intensity of the trapping lasers.

An alternative cooling technique is cold damping quantum feedback. Using the theory of Ref.~\cite{Genes2008} we have evaluated the minimum achievable mean phonon occupation number ignoring all sources of noise. In an extreme bad cavity limit ($\kappa=1000~\omega_m$), but otherwise using the system numbers quoted above, along with pump laser power of 100 mW and a feedback bandwidth $\omega_{\text{fb}} = 3~\omega_m$, we find $\langle n \rangle_{\text{min}} \simeq 0.56$. While this number lies within the quantum limit, we find that for our system numbers, cold damping is only effective deep within the bad cavity limit.

Future work will include the extension of this proposal to a three-mirror geometry, as well as the coupling of the levitated mirror to ultracold atomic and molecular systems, either for the quantum control of the state of the mirror, or conversely for the manipulation of the atoms. In particular, the generation, detection and control of non-classical motional states of the mirror will be considered. In addition, we will carry out a more detailed analysis of the optical coupling of the optical tweezers to the multilayered moving mirror.

We acknowledge stimulating discussions with K. Visscher, M. Aspelmeyer, G. Cole, K. Schwab and M. Vengalattore. This work is supported by the US Office of Naval Research, the US National Science Foundation and the US Army Research Office. GAP was supported in part by the University of Arizona NASA Space Grant.

\end{document}